\documentclass{INTERSPEECH2023}
\usepackage{multirow}
\usepackage{comment}
\usepackage{hyperref}
\usepackage{makecell}

\usepackage[belowskip=0pt,aboveskip=1pt]{caption}

\interspeechcameraready

\title{On the Robustness of Arabic Speech Dialect Identification }

\name{Peter Sullivan$^1$, AbdelRahim Elmadany$^1$, Muhammad Abdul-Mageed\textsuperscript{1,2}}

\address{
  $^1$The University of British Columbia, Canada\\
  $^2$Mohamed bin Zayed University of Artificial Intelligence, UAE 
  }
\email{\{prsull@student.,a.elmadany@,muhammad.mageed@\}ubc.ca}

\begin{document}

\maketitle
\begin{abstract}
Arabic dialect identification (ADI) tools are an important part of the large-scale data collection pipelines necessary for training speech recognition models. As these pipelines 
require application of ADI tools to potentially out-of-domain data, we aim to investigate how vulnerable the tools may be to this domain shift. With self-supervised learning (SSL) models as a starting point, we evaluate transfer learning and direct classification from SSL features. We undertake our evaluation under rich conditions, with a goal to develop ADI systems from pretrained models and ultimately evaluate performance on newly collected data. In order to understand what factors contribute to model decisions, we carry out a careful human study of a subset of our data. Our analysis confirms that domain shift is a major challenge for ADI models. We also find that while self-training does alleviate this challenges, it may be insufficient for realistic conditions.
\end{abstract}
\noindent\textbf{Index Terms}: Arabic speech processing, language identification, domain shift, dialect identification, Arabic language processing.

\section{Introduction}

Arabic faces significant challenges from a spoken language processing perspective. Mixing of dialectal Arabic (DA) and Modern Standard Arabic (MSA) in everyday speech~\cite{chowdhury2020effects}, means that performance of MSA-trained ASR models in realistic settings is deemed to be limited by weak performance of these models on DA. Similarly, purely DA ASR models are hampered by availability of only limited resources and lack of dialect identification (DID) tools that can aid creation of new dialectal datasets. Another challenge that impacts all language identification (LID) and DID tasks is the problem of performance on data from domains unseen during training. For the latter challenge, unsupervised domain adaptation methods have been proposed~\cite{duroselle2020unsupervised,nercessian2016approaches}.

One method that could potentially quickly boost DID models in absence of limited labeled data is speech self-supervised learning (SSL)~\cite{hsu2021hubert,baevski2020wav2vec}.
It is not sufficiently clear, however, how exactly SSL can fare  on realistic and out-of-domain settings. Similarly, the SSL representations potentially provide an alternative to x-vector segment representations, without overfitting to target domains. This is the case since output layers of models such as HuBERT have strong phonetic encoding~\cite{yang2021superb}, which could simulate earlier phonotactic approaches to language classification~\cite{singer2012mitll,abad2011l2f}.  
    
    In this work, we tackle the problem of robustness of Arabic speech DID, making the following contributions: (1) we quantify the performance of transfer learning from SSL, starting models under two settings: purely finetuned and self-trained; (2) we investigate use of SSL representations as baseline alternatives to i/x-vectors; (3) we assess performance on newly collected data to probe the limits of our transfer learning methods in a realistic data pipeline. 
\begin{figure}[]
\caption{Extracting Arabic utterances from the YouTube City data pipeline.}
\label{fig:pipeline} 
  \centering
  \includegraphics[width=\linewidth]{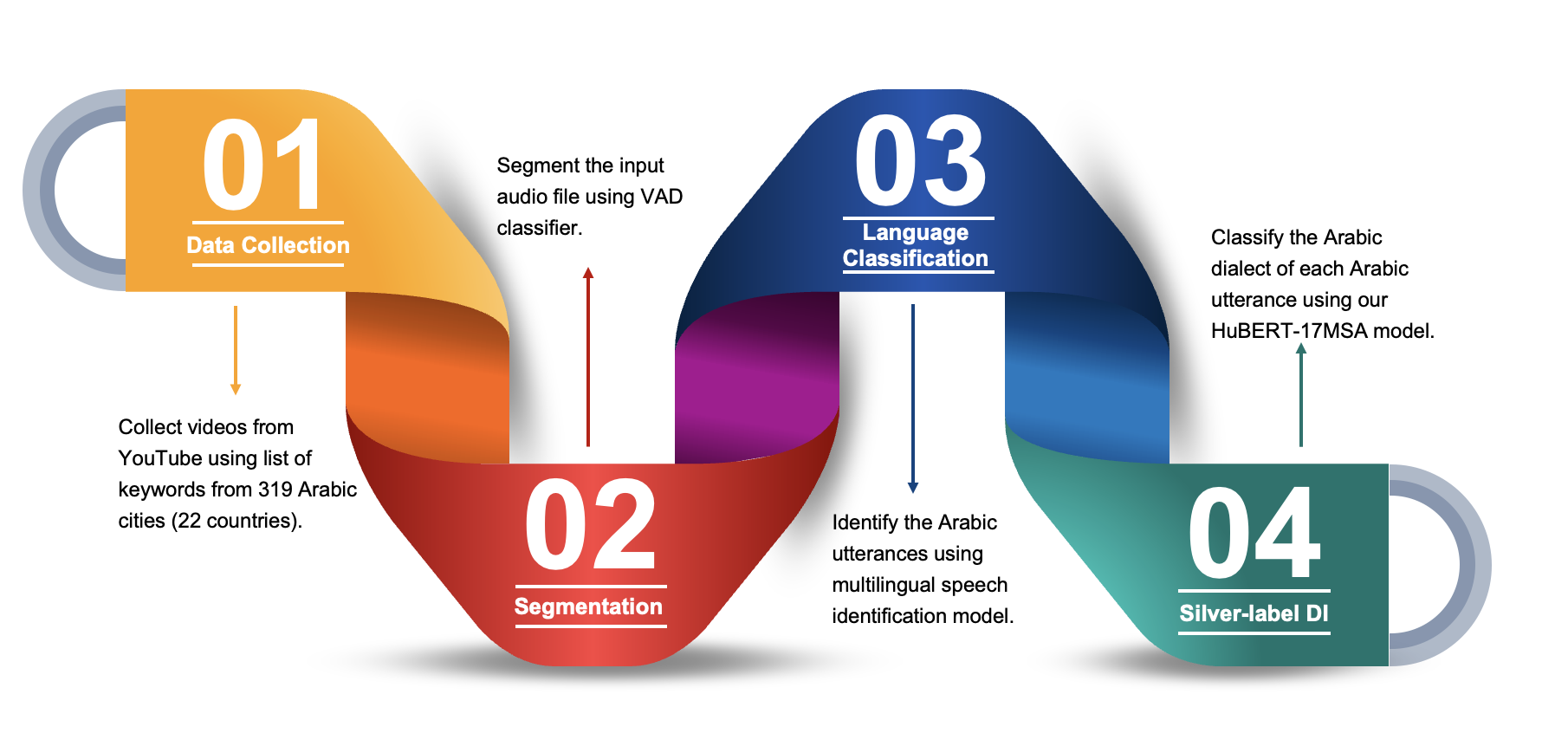}

\end{figure}

\section{Related work}

Several recent works considered ADI  ~\cite{ali2017speech, ali2019mgb,sadjadi20182017,chowdhury2020does,lin2020transformer}. As early as the 2015
NIST's Language Recognition Evaluation (LRE)~\cite{martin2015nist,sadjadi20182017}, the growing importance of neural architectures became apparent~\cite{lee20162015}. These include deep bottleneck features (DBN)~\cite{lee20162015,plchot2016bat}, DNNs for classification~\cite{plchot2016bat,yu2016utd}, i-vector post processing~\cite{lee20162015}, as well as summarization~\cite{plchot2016bat}. 
The 2017 challenge~\cite{sadjadi20182017}  further demonstrated the importance of neural embedding x-vectors for the classification task~\cite{snyder2018spoken}, marking the dominance of neural architectures over traditional, GMM-based approaches. 

The ADI-5 challenge, a coarse grain dialect and MSA classification task~\cite{ali2017speech}, similarly illustrates this transition. The winning system in the challenge used a siamese-network of CNNs to further post-process i-vectors~\cite{shon2017qcri}, and the second best system used a generative adversarial network in addition to a traditional Gaussian classifier for classification~\cite{bulut2017utd}. This trend towards more neural architectures continues with ADI-17, where neural end-to-end (E2E) architectures outperformed conventional i-vector approaches~\cite{ali2019mgb} with the best competition performance being an E2E ResNet architecture that employed novel pooling layers~\cite{cai2018insights} Deep learning methods used on these datasets include CNNs~\cite{chowdhury2020does,lin2020transformer} as well as transformers~\cite{lin2020transformer}.

SSL models, yet unexplored in ADI context, consist of a self-supervised pre-training objective such as a contrastive task~\cite{baevski2020wav2vec}, target prediction~\cite{hsu2021hubert}, or an autoregressive prediction~\cite{chung2020improved}. These models learn strong representations of linguistic features~\cite{yang2021superb}, which allow them to be finetuned on a variety of speech processing tasks. SSL models have already shown strong performance in LID~\cite{babu2021xls}, which suggests a strong potential of these models for DID. To the best of our knowledge, however, no studies have explored how best to use SSL models on ADI especially under potential domain shift conditions.

\section{Datasets and Preprocessing}\label{sec:datasets}
\subsection{Datasets}
In this work, we use three public datasets (ADI-5, ADI-17, MGB-2) alongside two datasets we newly collect (YouTube City, and YouTube Dramas). We introduce these next. \\ 
\begin{table}[ht]
\caption{The distribution of datasets in hour}
\label{tab:datasets}
\centering
\resizebox{0.8\columnwidth}{!}{%
\begin{tabular}{lrrrr} 
\toprule
 \textbf{Dataset} & \textbf{Dialects} & \textbf{TRAIN} & \textbf{DEV} & \textbf{TEST} \\
\midrule
\textbf{ADI-5}  & $5$ & $53.6$ & $10$ & $10.1$ \\
\textbf{ADI-17}  & $17$ & $3,033.4$  & $24.9$& $33.1$ \\ 
\midrule
\textbf{ADI-17 + MSA}  & $18$ &  $4,233.4$  & $26.9$ & $35$ \\
\textbf{YouTube City}  & $17$ &  $3,539$ &  --- & --- \\ 
\textbf{YouTube Dramas}  & $7$ &---& --- & $24.8$ \\ 
 \bottomrule

\end{tabular}%
}

\end{table}

\noindent\textbf{ADI-5}: A broadcast news dialectal identification corpus that is part of the MGB-3 Challenge~\cite{ali2017speech}. It consists of $\sim 74$ hours of audio segments labeled with broad categories of spoken dialects from the set \textit{\{ MSA, Gulf, Levantine, North African, Egyptian\}}. \\
\noindent\textbf{ADI-17}: Created as part of the MGB challenges (MGB-5)~\cite{shon2020adi17}, and consists of $\sim3,000$ hours of audio segments from YouTube programs covering a variety of different genres. The segments are split into $17$ different dialectal categories, allowing for much finer grain dialectal analysis than the ADI-5 corpus (although lacking MSA as a category).\\
\noindent\textbf{ADI-17 + MSA}: to partially alleviate the lack of MSA in ADI-17, we create a new dataset by supplementing ADI-17 training data with MGB-2  training split (which is mainly MSA~\cite{ali2016mgb}) and the MSA part of ADI-5 training split. For the development dataset, we concatenate ADI-17's development set with the MSA portion of ADI-5's development set.\\
\noindent\textbf{YouTube City}: we randomly pick up $91,365$ videos ($\sim8,746$ hours) from an in-house massive audio dataset collected from YouTube using a list of keywords that covers $319$ Arabic cities from $22$ countries. It consists of $238.2$K videos ($\sim62$k hours). We apply a predefined preprocessing pipeline on the selected data to extract the Arabic utterances (Section~\ref{sec:pipeline} for more details). \\
\noindent\textbf{YouTube Dramas}: we manually collect Arabic dialectical drama series from YouTube that cover seven dialects: \textit{Algerian}, \textit{Egyptian}, \textit{Emirati}, \textit{Jordanian}, \textit{Moroccan}, \textit{Palestinian}, and \textit{Yemeni}. Similar to the YouTube City data, we apply a predefined preprocessing pipeline to extract the Arabic utterances. The final YouTube Dramas dataset totals $24.8$ hours extracted from three series for each dialect. 

We ensure no overlap between ADI-17,  the YouTube City, and YouTube Dramas datasets.  Table~\ref{tab:datasets} shows the distribution of the datasets.

\subsection{Preprocessing Pipeline}\label{sec:pipeline}

As shown in Figure~\ref{fig:pipeline}, the preprocessing pipeline for extracting the Arabic utterances from YouTube City collected data relies on the following four steps:

\noindent\textbf{(1) Data Collection}. Inspired by~\cite{abdul-mageed-etal-2020-toward}, we create a set of search keywords using a list of cities concatenated with country name, and use this to select videos. As stated earlier, the collected data consists of $238.2$K videos ($62$k hours).  For the self-training in this paper, we randomly select a maximum of $6,500$ videos for each of the countries in ADI-17, leaving us with $91.4$K videos ($8.8$K hours).\\  
\noindent\textbf{(2) Segmentation}. We use the voice activity detection model \textit{pyannote.audio}\footnote{\href{https://huggingface.co/pyannote/voice-activity-detection}{https://huggingface.co/pyannote/voice-activity-detection}}~\cite{Bredin2020, Bredin2021} to segment the audio files, keeping only utterances with length more than $3$ seconds.  The model give us  $2$M utterances from the $91.4$K files.\\
\noindent\textbf{(3) Language Identification}. We use the multilingual speech identification model \textit{lang-id-voxlingua107-ecapa}\footnote{\href{https://huggingface.co/speechbrain/lang-id-voxlingua107-ecapa}{https://huggingface.co/speechbrain/lang-id-voxlingua107-ecapa}}~\cite{speechbrain, valk2021slt} to extract Arabic utterances only from the data. The model is able to detect $1.7$M Arabic utterances from the $2$M utterances.\\
\noindent\textbf{(4) Silver-Label Dialect Identification}. We create predictions for each utterance of the Arabic utterances, sorting theses into four groups for self-training: \textit{Surrogate labelled}, labeled using the country of origin; \textit{High}, \textit{Medium}, and \textit{Low} confidence, acquired by splitting the data based on the prediction confidence from  HuBERT-17MSA classifier (see Section~\ref{subsec:ft_models}). The $442.6$K utterances ($26.58$\%) of the Arabic data are considered `correct' (with weak labels) regarding the matching between the original country name and the model prediction label. (Section~\ref{subsec:self-training} for more details).

\section{Methods}

\subsection{Finetuned SSL Models}\label{subsec:ft_models}
To evaluate SSL models, we choose \textbf{(i)} the massively multilingual model XLS-R 300m~\cite{babu2021xls} as our example of the wav2vec 2.0 model, and pick \textbf{(ii)} HuBERT-large-ll60k~\cite{hsu2021hubert} as a contrasting monolingual English HuBERT model. Evidence from SUPERB indicates that while HuBERT may not be trained on Arabic, it has strong phonetic modeling capabilities~\cite{yang2021superb}.

We aim to deploy these pretrained models as dialect ID systems by adding modified TAP layers~\cite{cai2018exploring}. We first add a projection layer to reduce the dimensionality of our output, then mean pool the projected outputs over the temporal dimension before we pass this through a final classification layer.

Because these pretrained models are designed for use primarily in ASR, the finetuning process must be adjusted for DID. We do so by exploring the configuration space through a random search~\cite{bergstra2012random}, which enables for an efficient search due to the likelihood of many hyperparameters not mattering significantly. Using a fixed compute budget for our two fine-tuned models of one node month of compute time (30 iterations) each, we search the following training hyperparameters: \textit{Batch Size}, the number of samples per each gradient update;  \textit{Freeze Steps}, the number of updates with only the final prediction layer thawed; \textit{Learning Rate}, a variable multiplying the size of the gradient update; \textit{Max Steps}, the total number of steps in our training;  \textit{Sample Duration}, the duration of the random sample taken from the source audio;  \textit{Thaw Depth}, which layers of the model to thaw during fine-tuning. Additionally, we experiment with applying Layer-Norm and Attention (LNA) fine-tuning, which is a fine-tuning technique that freezes all layers except for layer norm and multihead self-attention~\cite{li2020multilingual}. We summarize the ranges and distributions used for this finetuning exploration in Table \ref{tab:hyperparams}. We adopt the original wav2vec 2.0 tri-state learning rate schedule, which consists of breaking the total training steps into a $10$\% ramp up to max learning rate, $50$\% plateu, and $40$\% cooldown period.

We directly fine-tune the xls-r-300m and HuBERT-large-ll60k models on our ADI-17 training set, to create fine-grained ADI models \textit{xls-r-300m-17} and \textit{HuBERT-17}. Since we realistically also want to differentiate between MSA and DA, we train a HuBERT model on the ADI-17MSA dataset to create \textit{HuBERT-17MSA}.

\subsection{SSL Features}
While both HuBERT and wav2vec 2.0 create outputs that can act as strong representation of the audio input, DID tasks can be prone to models overfitting to channel information or other aspects of the audio that are non-linguistically relevant~\cite{bovril2012arabic}. This motivates us to acquire representations  using the psuedo-labels from the k-means clustering of HuBERT output features. 

We select representations from the Base model layer with the best Phone-Purity and PNMI, which occurs as output from layer $10$~\cite{hsu2021hubert}~\footnote{While the main HuBERT paper uses layer $9$ for training of the HuBERT Large and XL, their graphs indicate potentially better attributes from using layer $10$ with respect to PNMI and Phone-Purity~\cite{hsu2021hubert}}. Our process for generating the pseudo-labels is as follows: first, we exclude very short and very long utterances by selecting ADI-17 train files that are between $5$ and $30$ seconds long. We then pass these raw files through the HuBERT model, extracting the feature representations from layer $10$. Because what level of phonetic information captured is related to the number of clusters, we treat this as a hyperparameter in our experiment: using a subsample of $10$\% of the resulting frames, we train $5$ different k-means models using a choice from the set \textit{\{$200$, $400$, $600$, $800$, $1000$\}}. Then, we assign cluster labels to the extracted features for our data. Because HuBERT covers $320$ samples of audio per each prediction frame, for 16khz sample rate audio this leaves us with sequences of between $250$ and $1500$ labels. We  compress these sequences into a length normalized representation essentially corresponding to the unigram count of each label divided by the total length of the sequence.

For classification, we pass the length normalized representations through a single layer neural network, training with cross-entropy loss.

\begin{table}[!htbp]
\centering
\caption[HuBERT Fixed representation hyperparameters]{Overview of hyperparameter search space for our pseudo-transcript classification models. Optimal results, batch: $256$; clusters: $1000$; learning rate: $1 \times 10^{-2}$.
}
\label{tab:percept_hyper}
\resizebox{0.8\columnwidth}{!}{%
\begin{tabular}{ll} \toprule 
 \textbf{Hyperparameter}  & \textbf{Range} \\
 \midrule
     Batch Size  & $[64,128,256,512]$ \\ 
    \textit{k}-means Clusters  & $[200,400,600,800,1000]$ \\ 
 Learning Rate & $[1\times 10^{-2},1 \times 10 ^{-3},1 \times 10 ^{-4}]$ \\ 

 \bottomrule 
\end{tabular}%
}

\end{table}
\vspace{0.1cm}

\subsection{Self-Training}\label{subsec:self-training}

For self-training, we use silver labeled data from our YouTube City dataset. We group the newly collected data into several clusters based on (1) YouTube channel country information and (2) label confidence. We hypothesize that channels which are (in)consistently classified as being from the Gulf region are likely sources of MSA audio in contexts unfamiliar to the model. Similarly, we split up the silver labeled data into \textit{low}, \textit{medium}, and \textit{high} confidence predictions. These categories are then concatenated to the ADI17+MSA dataset independently.

\begin{table}[!htbp]

\caption[Self-training settings]{ Breakdown of how the YouTube City data was split for self-training. Confidence intervals were based on 33 percentile intervals of the HuBERT-17 predictions.
}
\label{tab:self_train_break}
\centering
\resizebox{0.95\columnwidth}{!}{%
\begin{tabular}{ll} \toprule 
 \textbf{Self training settings}  & \textbf{Description} \\
 \midrule
\multirow{1}{*}{Surrogate label} &  Label via country of origin \\
    Low confidence  & $<54.24\%$  confidence \\ 
 Medium confidence & $54.24\% \leq x <87.84\% $ confidence \\
 High confidence & $ >87.84\% $ confidence \\

 \toprule 
\end{tabular}%
}
\end{table}

\begin{table}[]

\caption[Finetuning hyperparameters]{
The search space for our random search of finetuning hyperparameters as well as optimal configurations on ADI-17 \textit{Dev}. We set the batch size such that a total of $75$ seconds of audio can comfortably fit onto each V100 GPU, of which we train with $4$ at a time. All values are picked from uniform distributions except for the learning rate (log uniform).
}
\label{tab:hyperparams}
\centering
\resizebox{0.95\columnwidth}{!}{%
\begin{tabular}{lcll} \toprule 
 \textbf{}  & \textbf{Range} & \textbf{HuBERT} & \textbf{xls-r} \\
 \midrule
     Batch Size  & $4\times \lfloor\frac{75}{Duration} \rfloor$  & $16$ & $9$\\ 
 Freeze Steps  & $[0,1000]$ & $192$ & $960$\\

 Learning Rate  & $[1\times 10^{-5},1 \times 10 ^{-2}]$ &$6 \times 10^{-4}$ & $1.2 \times 10^{-3}$ \\
  LNA  & $[$True, False$]$ & False & True\\ 

 Max Steps & $[20k,40k]$ & $29225$ & $35956$\\ 
  Duration  & $[4,18]$ seconds & $4.69$ & $8.33$\\ 

 Thaw Depth  & $[0,23]$ & $3$ & $4$\\ 

 \bottomrule 
\end{tabular}%
}
\end{table}
\vspace{0.1cm}

\section{Experiments}
\textbf{ADI-17.} We evaluate our xls-r-300m-17, HuBERT-17, and HuBERT features models on the ADI-17 Test set to understand performance of these methods on in-domain classification.\\
\textbf{ADI-17 Transfer to ADI-5.} By pooling the country labels into the corresponding coarse region labels of ADI-5,\footnote{Gulf: KSA, UAE, OMA, IRQ, KUW, YEM, QAT; Levantine: LEB, PAL, JOR, SYR; Egypt: EGY, SUD; North Africa: MOR, MAU, LIB, ALG} we test the out-of-domain performance of our models trained on ADI-17 on the ADI-5 Test set.\\
\textbf{Domain Shift.} We similarly, examine whether self-training on large diverse audio aids performance on a different domain distribution in comparison to models simply finetuned on ADI-17. We use our YouTube Dramas dataset as our Test set. This set consists only of drama series audio, unlike the multi-genre ADI-17 and YouTube City corpora, and may reflect a significant style of data the model is tailored for.
For training, we use our existing HuBERT-17MSA checkpoint as well as the HuBERT-large-ll60k checkpoint as starting points, and use the identified finetuning hyperparameters from our HuBERT-17 training. We report results in macro-$F_1$ over the target classes.\\
\noindent \textbf{Human Analysis.}
To understand the quality of the self-training labels, we randomly pick $25$ utterances from the set of utterances where HuBERT-17MSA prediction and identified origin country do not match. We do this for four of the dialects: Emirati, Jordanian, Moroccan, and Sudanese. Then, a native speaker expert manually listens to and annotate the selected utterances. We check whether the utterance belongs to the identified country, whether it is DA or MSA, and whether or not the HuBERT-17MSA prediction is  correct.


\section{Results}
\begin{table}[]

\caption[ADI-17 Results]{Models' results on ADI-17's development and test datasets in term of macro $F_1$ score. \textsuperscript{$\bf\dagger$}Our runs. \textsuperscript{$\bf\star$}as reported in~\cite{shon2020adi17} and  \textsuperscript{$\bf\ddag$}as reported in~\cite{ali2019mgb}, we do not have dev $F_1$ scores for these runs. \textbf{Bold:} the best results show in boldface.}
\label{tab:adi_res}
\centering
\resizebox{0.9\columnwidth}{!}{%
\begin{tabular}{llrr} 
\toprule 
\textbf{Type} & \textbf{Model} & \multicolumn{1}{c}{\textbf{ADI-17 Dev}} & \multicolumn{1}{c}{\textbf{ADI-17 Test}} \\
                 \midrule
\multirow{4}{*}{Fixed} & Majority Class & $1.33$  & $0.67$ \\
 & HuBERT Features\textsuperscript{$\bf\dagger $} &  $52.07$ & $51.36$ \\
& i-Vector\textsuperscript{$\bf\star$}  &  ---   & $60.60$ \\
& x-Vector\textsuperscript{$\bf\star$}  &  ---   &  $72.40$  \\
                \hline
\multirow{3}{*}{E2E}  &  ResNet\textsuperscript{$\bf\star$} &   ---  & $82.69$  \\
&  DKU ResNet\textsuperscript{$\bf\ddag$} &   $97.4$  & $\bf94.9$  \\
                &  HuBERT-17\textsuperscript{$\bf\dagger$} &	$92.23$   & $92.12$   \\
         &    xls-r-300m-17\textsuperscript{$\bf\dagger$}  &	$90.77$  &  $90.20$ \\ \bottomrule 
                
\end{tabular}%
}
\end{table}

\begin{table}[h]

\caption[ADI-5 Results]{Results of ADI-17 trained models ADI-5 Test (minus MSA segments).  We do not compare prior submissions to the ADI-5 challenge as these predict on the full set.  $^\star$F\textsubscript{1}  $45.01$ on full ADI-5 with MSA}
\label{tab:adi5}
\centering
\resizebox{0.5\columnwidth}{!}{%
\begin{tabular}{lr} \toprule 
          \textbf{Model}         & \textbf{F\textsubscript{1}} \\
                 \midrule Majority Class  & $10.92$
 \\
HuBERT Features  & $67.71$ \\
 HuBERT-17   &  $\bf80.36$ \\
 XLS-R-300m E2E   & $76.20$ \\ 
HuBERT-17MSA$^\star$  & $37.17$ \\\bottomrule
\end{tabular}%
}
\end{table}

\begin{table}[]

\caption{Results on ADI17+MSA's development dataset and YouTube Dramas' test dataset in the term of macro $F_1$ score over target classes. Setting refers to the self-training starting checkpoint: HuBERT-large-ll60k (\textit{Vanilla}) or our fine-tuned HuBERT-17MSA (\textit{17MSA}). \textsuperscript{$\star$}Self-training with the surrogate labels from scratch failed to converge.}
\label{tab:youtube}
\centering
\resizebox{0.9\columnwidth}{!}{%
\begin{tabular}{clcc} \toprule  
\textbf{Setting} & \textbf{Model} & \makecell{\textbf{ADI17+MSA Dev} $\uparrow$ \\ (18 classes)} & \makecell{\textbf{Dramas Test} $\uparrow$ \\ (7 classes)}  \\
                 \midrule 
 & HuBERT-17   &  - & $0.91$  \\
 & HuBERT-17MSA   & $\bf89.39$ & $5.09$ \\
 \hline
\multirow{4}{*}{\rotatebox[origin=c]{90}{17MSA}} & HuBERT-YT\textsubscript{s}   & $88.41$ & $43.12$ \\
& HuBERT-YT\textsubscript{h}   & $87.76$ & $40.01$ \\
& HuBERT-YT\textsubscript{m}   & $88.82$ & $39.40$ \\
& HuBERT-YT\textsubscript{l}  & $88.94$ & $43.19$ \\
 \hline
\multirow{4}{*}{\rotatebox[origin=c]{90}{Vanilla}}  &  HuBERT-YT\textsubscript{s}   & $\star$ & $\star$ \\
 & HuBERT-YT\textsubscript{h}   & $88.89$ & $\bf45.10$ \\
 & HuBERT-YT\textsubscript{m}   & $88.51$  & $42.89$ \\
 & HuBERT-YT\textsubscript{l}  & $88.62$ & $43.20$ \\
 \bottomrule 
\end{tabular}%
}
\end{table}

\noindent\textbf{ADI-17 Results.}
Compared to xls-r-300m, under our limited tuning HuBERT does surprisingly well. This is the case despite the model having no exposure to Arabic during pre-training (see Table \ref{tab:adi_res}). This performance can potentially be explained by a HuBERT better phonetic modelling compared to wav2vec 2.0 based training~\cite{yang2021superb}. Using SSL models as feature extractors for simple classification appears to work as a reasonable baseline. Although not as competitive as even i-vector or x-vector systems, the simplicity of the approach could strike a balance between capturing the strength of SSL models and the potential for interpretability.

\noindent\textbf{ADI-17 Transfer to ADI-5.}
Excluding MSA predictions, our HuBERT model finetuned on ADI-17 performs well. This is the case despite a significant drop in performance (see Table \ref{tab:adi5}), likely due to difference in channel and domain (e.g., topics covered and their distribution) of the ADI-5 dataset in comparison to the YouTube data for ADI-17. This hints at the potential problems in applying ADI models in the wild. Our SSL features method appears to be fairly robust to these same domain differences, and improves in relative performance on the ADI-5 (minus MSA) set.

Training on MSA appears to have a large detriment to performance on the DA portions of the ADI-5 dataset. Since this model has seen the MSA parts of the ADI-5 train set, it is likely that is has used channel differences between the MSA (broadcast television) and ADI-17 DA (YouTube) as a shortcut for classification. This illustrates the need for MSA that matches both domain and recording characteristics of DA, a potentially challenging issue due to sociolinguistic usage differences between MSA and DA.

\noindent\textbf{Domain Shift Performance.}
The YouTube Dramas dataset appears to be a difficult domain for all the models (see Table \ref{tab:youtube}), with models that have only been finetuned appearing to fair the worst. Self-training, in comparison, alleviates some of these issues. Considering performing self-training from scratch vs. an existing finetuned checkpoint, we observe a benefit of training from scratch. However, the finetuned models are able to achieve nearly the same results by training for only a few steps with the augmented dataset ($500$-$1000$ steps vs. $28500$-$29000$ steps) and this short-training phenomenon likely needs further exploration. 

\noindent\textbf{Human Analysis.}
We find that $74.71$\% of the surrogate labels do not belong to the country identified through the collection process, but that of these mislabeled utterances the model is able to detect the correct dialect of $11.49$\%. In addition, the model is not able to recognize $85.87$\% of the MSA utterances, reflecting the fact that MSA on different topics or recording conditions from the training data may be difficult to detect. 



\section{Discussion}
 Quantized features from SSL models appear to be sufficient for LID and DID. These, however, could potentially be improved by either n-gram representations of features or different classification methods (e.g. CNN) that can capture a larger temporal window.

For them to be realistic and fit for real-world use, LID and DID benchmarks should consider either zero-shot or few-shot out-of-domain classification tasks. Similarly, the performance gap in detecting MSA utterances indicates the need for diverse MSA training corpora that cover both diverse recording conditions as well as a wide host of topics.

 In addition, while using a small hyperparameter search space appears to be sufficient for reasonable performance for fine-tuning DID models from pretrained SSL models, there is likely room for further optimization by expanding the number of iterations as well from increasing the maximum training time. 
 
YouTube audio may reflect vastly different recording and acoustic conditions, which we assume could  be random in nature at times. The exact distribution of these characteristics could be further explored with respect to type of content and country of origin.

\section{Conclusions}
We presented experiments quantifying out-of-domain (both channel differences and topic shift) performance of Arabic DID models trained using self-training, finetuning, and a fixed representation approach. Our results emphasize the difficulties of domain shift for DID models and the importance of evaluating models on realistic settings.

We identify a number of directions for future work, including exploration of accented DID which would allow a better understanding of how these models may rely on phonotactic elements; evaluation of few shots settings for out-of-domain DID; and improvements to SSL representation models. Concretely, we also identify a strong need for diverse (formality, genre, channel, accent) MSA datasets.

\bibliographystyle{IEEEtran}
\bibliography{mybib}

\end{document}